**Efficiency versus effort: a better way to compare best photovoltaic research cell efficiencies?**

Phillip J. Dale[1] and Michael A. Scarpulla[2]

1. Department of Physics and Materials Science, University of Luxembourg, Belvaux, Luxembourg

2. MSE and ECE Departments, University of Utah, Salt Lake City, Utah, USA

Corresponding authors: phillip.dale@uni.lu and mike.scarpulla@utah.edu

**Abstract**

Frequently, trends in record AM1.5 power-conversion efficiencies versus time, such as the NREL efficiency chart, are used to analyze the relative merits of different photovoltaic material technologies. However, this approach belies the effort expended in achieving these levels of performance. We introduce cumulative publications as a proxy for total R&D efforts and find surprisingly that silicon, Cu(In,Ga)Se$_2$ (CIGSe), CdTe, and halide perovskite technologies have each followed essentially the same learning curve of 20-24% efficiency within 10,000 publications and a consistent marginal rate of 5% efficiency increase per factor of 10 in publications. While learning spillover from non-PV technologies, cross-pollination from other PV technologies, and hidden commercial effort are not accounted for by this metric, this analysis still yields useful and novel insights into PV technology trajectories. Trajectories below this learning curve have required more total effort per performance and plateaus of efficiency stagnation at large numbers of publications may indicate (but do not guarantee) the existence of fundamental barriers to commercially relevant performance. Lastly, examples to watch are identified for technologies currently exhibiting higher marginal slopes, including some that appeared dormant by this metric in past years.

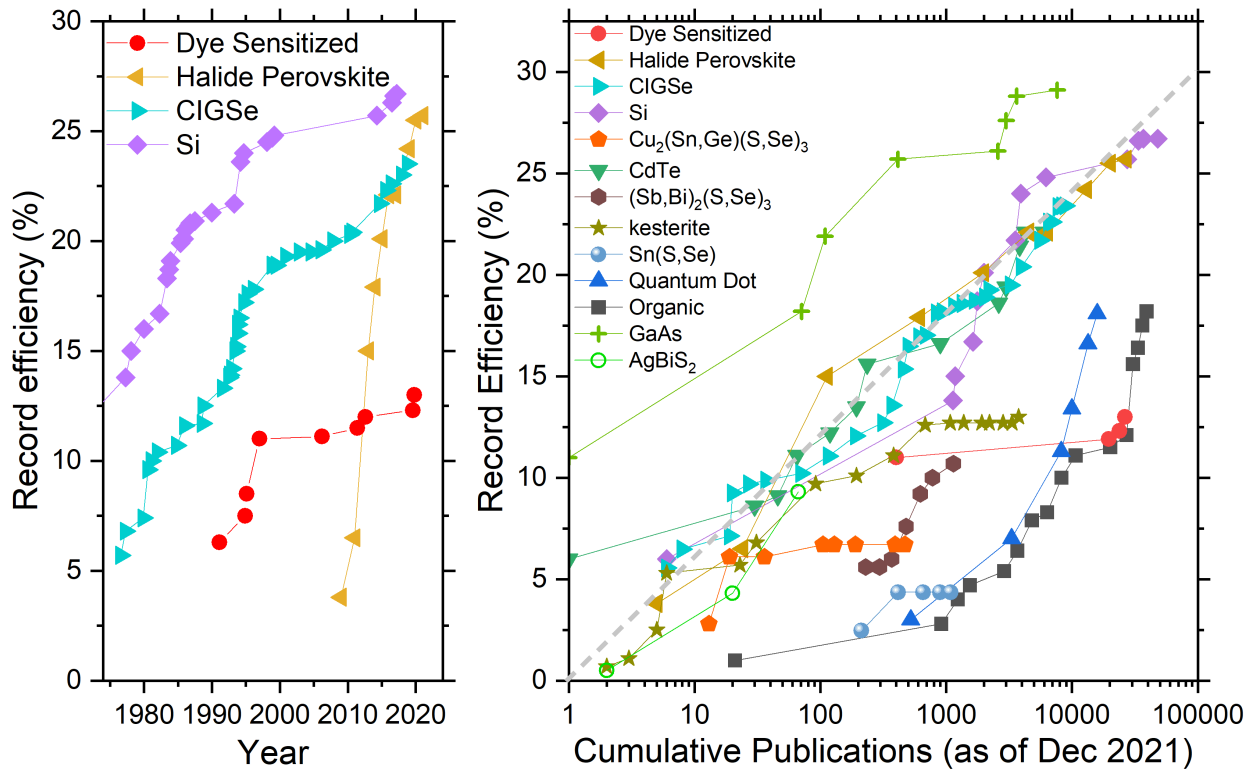

Figure 1 – (a) Record certified efficiency vs. time for four selected single-junction PV technologies exhibiting apparently very different trajectories [1]. (b) Record efficiencies for a larger set of technologies as a function of cumulative publications since the first working solar cell was announced. The dashed line indicates a learning slope of 5% absolute efficiency per each factor of 10 in publications, taken as a proxy for effort.

For decades, progress in different photovoltaic (PV) technologies has been tracked by NREL on a chart of record cell efficiency versus date [1]. Researchers and technologists have used the relative position and trajectory of different material technologies in deciding whether to change or add new technologies to their portfolio. Figure 1(a) shows the record certified AM1.5 power conversion efficiency as a function of year of achievement for four different technologies. Silicon solar cells have recently achieved a record efficiency of 26.7% which is very close to their expected limit near 29% due to Auger recombination [2], with the rate of efficiency improvement slowing as it nears this limit. The lower current efficiencies of $Cu(In,Ga)(S,Se)_2$ (CIGSe) and dye sensitized solar cell (DSSC) technologies could be rationalized by proponents as due to the simple fact that research began later; given more time they might reach the same performance as Si has after nearly 70 years. Additionally, Si has benefitted from know-how in the electronics industry, so these PV-only material technologies are arguably making better progress. Halide perovskites have nearly achieved their theoretical maximum efficiency in just over a decade. These observations raise the question: "How can intrinsically-better-suited PV material technologies relevant for commercialization and real-world impact be recognized, and recognized early?".

We first noted that making inferences about the intrinsic qualities of a PV material technology from the NREL efficiency-time chart could be misleading because there is no accounting for the number of researchers working in the field or their resources. For example, CdTe solar cells were worked on only by a few dozen groups in the decades from the first polycrytalline thin film cell in 1962 until the early 2000's, whereas organic, DSSC, quantum dot, and halide perovskite technologies have attracted hundreds or even thousands of groups. As of today, CdTe has over 30 $GW_p$ of modules installed in utility-scale installations due to huge capital investment and commercialization while the others aforementioned are still primarily in the research stage. Is there any way to predict if similar capital investment and commercial effort could bring other technologies to scales relevant to global energy production?

We began from the premise that superior PV material technologies should exhibit higher performance per effort expended – good materials should make it easy. Unfortunately, historical worldwide data on funds or person-hours put towards each material technology are not available. Thus, we experimented with using the cumulative number of technical publications as a proxy for R&D effort since academic researchers report their findings and know-how is cumulative.

We have made the simplistic assumption that on average each publication advances the field a similar amount. This assumption comes with the proviso that publication trends such as frequency [3][4] and information content [5] may change with time, and are still debated. Both breakthrough single papers and incremental advances over many papers result in increased efficiency; many technologies on the NREL chart exhibit slow plateaus punctuated by steps. Examples at different efficiency levels include Na-glass for CIGSe, Se alloying for CdTe, non-fullerene acceptors for organics, and external radiative efficiency optimization for GaAs. Thus, a more nuanced view is that the total probabilities both of making breakthroughs and of making incremental advances scale with the number of researchers, researcher effectiveness, time spent, and resources available. All of these arguably also correlate with publication volume.

We point out three effects not captured in our analysis that would yield efficiency increases with lower publication rates (i) "hidden commercial effort" (ii) "spillover": advances and resources in technologies involving the same materials but for non-PV applications and (iii) "cross-pollination": PV technologies can benefit each other especially when they utilize similar materials, fabrication, techniques, or equipment and infrastructure.

Despite any imperfections of the methodology, Figure 1(b) enables a different perspective for comparisons of PV technologies which we hope will yield insights and spark further thoughtful analysis. First, we comment on a surprising finding for CdTe, CIGSe, and halide perovskites; these thin-film technologies have followed very similar efficiency vs. effort trajectories over more than three orders of magnitude in publications. These technologies all hit the marks of 10% efficiency within 100 publications and 20-24% in fewer than 10,000 publications and have in each decade of publications followed an average slope of approximately 5% absolute efficiency per factor of ten increase in publications. We refer to these as the mainline polycrystalline thin film technologies. These material technologies all have grain boundaries, heterointerfaces, and foreign substrates which could add non-radiative recombination. However, in each instance some combination of quirks of the materials' phase diagrams and or special thermochemical treatments have been found to conspire to passivate interfaces and grain boundaries, reducing recombination.

On plots of efficiency vs year (Figure 1(a)), contrast is often drawn between halide perovskites, which appear to improve quickly, and thin film CdTe/CIGSe technologies which follow diminishing returns. This difference has been claimed to indicate that halide perovskites possess superior attributes as PV materials. Indeed, halide perovskites have achieved higher efficiencies in a shorter time period, but this was accomplished within many times more publications (~27,000 publications compared to ~6,000 for CdTe and ~8,800 for CIGSSe). However, our analysis suggests that the average efficiency return on effort, at least by those motivated by publishing R&D results, are very similar for halide perovskites, CdTe and CIGSe. This implies that exponential growth vs. year is driven more by exponential growth in researchers and efforts expended than the technology being qualitatively different than these other high-efficiency polycrystalline thin film technologies. We hypothesize that, in addition to the excellent properties and performance of halide perovskites, lower barriers to entry in terms of simpler techniques, lower non-vacuum equipment capital costs, and existing tools and infrastructure from prior organic/QD/and DSSC PV and electronics research have played parts in attracting large numbers of researchers to this technology. Many of these effects may be related essentially to their lower synthesis temperatures, which in turn relates to intrinsic properties like bond strength. However, Figure 1b argues that the same total required work was simply accomplished in a shorter time frame by more researchers and or higher publication rates. This experience suggests that devoting more effort and resources to other reasonably-well-performing technologies may yield predictable returns, whether or not the technology appears stalled on an efficiency vs time chart. We note that the crossover of techniques and infrastructure into halide perovskite PV provides an argument for the value of continuous investment in fundamental research.

Next, we comment on Si and GaAs technologies that both benefit from know-how and resources related to multi-billion dollar, commercialized, non-PV technologies. Nonetheless, the trajectory of Si technology, perhaps coincidentally, does overlay quite closely with the mainline thin film PV group mentioned above. GaAs technology has undoubtedly benefitted from the development of III-V optoelectronics such as LEDs and lasers, especially given the sensitivity to water and oxygen which MBE and MOVPE were designed to eliminate. This, we believe, explains the fact that its data shares the same slope as the mainline technologies but with a sizable vertical offset. It is indisputable that the degree of perfection achieved in III-V based cells is a remarkable achievement. The fact that the slope of efficiency vs publications is very similar to that of the mainline thin film technologies invites speculation that there may be a universal learning curve for PV efficiency as function of R&D efforts for technologies capable of reaching commercially-relevant performance and that offsets on our chart can be mostly understood as arising from the use of publications as the proxy variable. Another speculation arises from the logarithmic dependence on publications; is this fundamentally a reflection of the logarithmic dependence of $V_{oc}$ on external radiative efficiency which in turn depends on elimination of non-radiative recombination and perfection of optical design?

Next we comment on alternative chalcogenide thin film technologies such as $Cu_2ZnSn(S,Se)_4$ (kesterite), $Cu_2(Sn,Ge)(S,Se)_3$, $Sn(S,Se)$ and $(Sb,Bi)_2(S,Se)_3$. In general, these fall below the mainline thin film group indicating that they may suffer from material problems such as lifetime-killing bulk or interface defects, phase segregation, band-tailing or ineffective doping. After an initially promising development kesterite performance has stagnated, as has $Cu_2(Sn,Ge)(S,Se)_3$ and $Sn(S,Se)$. However, $(Sb,Bi)_2(S,Se)_3$ is an interesting case, exhibiting a higher marginal slope than the mainline technologies while being relatively unexplored with just over 1,100 publications. This is certainly a technology to watch. A disadvantage for all these materials, unlike silicon and GaAs, is that they are mostly being investigated for PV, which limits spillover from other application areas. We note that, like halide perovskites capitalizing on prior experience and infrastructure in organic and dye-sensitized, especially kesterite benefitted from CIGSe.

Only time and stringent analysis of achievable property values will tell if these technologies will continue to improve or stagnate in terms of real-world impact.

Lastly, we turn with interest to comparisons amongst DSSC, organic and quantum dot PV. In our early versions of our analysis approximately a decade ago, DSSC was the clear winner and organic and quantum dot technologies looked to be on trajectories of diminishing returns – thousands upon thousands of publications with only small increments in efficiency. However, today the case is reversed. After an incredible volume of papers published, organic and quantum dots are now on trajectories with marginal rate above 5%. In terms of absolute efficiency organic and quantum dot PV are still below the lab-scale 20% efficiency mark which is approximately the requirement for primary electricity applications. We note that much of the recent QD records involve halide perovskite materials, a case of cross-pollination. The recent sea change in the trajectory of organic PV has been caused by true breakthroughs in non-fullerene acceptors; a case in point of the wisdom of continued fundamental research even when technology assessments would predict little gain.

In conclusion, using a simple metric for R&D effort, it appears that c-Si, CIGSSe, CdTe, halide perovskites, and to some extent GaAs PV technologies follow a similar learning curve. This may offer an effective method for assessing the potential of emerging PV materials in the R&D stage where cumulative production learning curves do not exist. An improvement to the metric could include elements related to patents, citations or total publication word count. Amongst technologies deviating from the main learning curve, organic and quantum dot technologies illustrate that breakthroughs may occur at any stage. Materials like $AgBiS_3$ are quickly gaining efficiency leveraging knowledge from the hybrid perovskite system [6]. However, if the goal is for research scientists to provide the materials and means for the planet to go 100% renewable in this generation with multi-junction photovoltaics then the decision to work on certain materials also depends on other factors. Having an efficiency greater than 20% is a prerequisite estimated to offset the fixed costs of installation and balance of materials; but research cell efficiency does not capture challenges to module production such as area scaling while maintaining uniformity and performance, cost-effective manufacturability at scale, supply chains for materials, 25 to 50 year stability, and high energy return on the energy invested.

Methods:

Efficiency data as a function of time were taken from the NREL website[1]. Efficiency data not in the NREL website was extracted directly from source publications. The number of publications in any given year was found by running a Web of Science [7] search using the associated technology names and the phrase (photovoltaic OR solar cell). For the year of any given technology record, this is matched to the cumulative number of publications of the technology since year of publication of the first device. Occasionally, a NOT statement is added to remove technology which is unrelated with that being searched for. For example, SnS and SnSe are written about a lot in kesterite literature since they are secondary phases in this system, but they are not being actively researched in the article for use as absorber layers by themselves. For Si, CdTe, and GaAs the Scopus database [8] was used to add publications pre-1975, and Google Scholar [9] for pre-1960 Si publications. A full list of search terms is given in the supplementary information.

Data Availability:

The data of figure 1 can be found at https://doi.org/10.5281/zenodo.7115026 hosted at Zenodo.org. The dataset will be updated on a yearly basis.

Acknowledgements:

We acknowledge Laurence Peter for helpful discussions and the referees for their constructive comments and suggestions.

Bibliography:

[1] Best Research-Cell Efficiency Chart, (n.d.). https://www.nrel.gov/pv/cell-efficiency.html (accessed September 12, 2022).
[2] R.M. Swanson, Approaching the 29% limit efficiency of silicon solar cells, in: Conference Record of the Thirty-First IEEE Photovoltaic Specialists Conference, 2005., 2005: pp. 889–894. https://doi.org/10.1109/PVSC.2005.1488274.
[3] L. Bornmann, R. Haunschild, R. Mutz, Growth rates of modern science: a latent piecewise growth curve approach to model publication numbers from established and new literature databases, Humanit Soc Sci Commun. 8 (2021) 1–15. https://doi.org/10.1057/s41599-021-00903-w.
[4] D. Fanelli, V. Larivière, Researchers' Individual Publication Rate Has Not Increased in a Century, PLOS ONE. 11 (2016) e0149504. https://doi.org/10.1371/journal.pone.0149504.
[5] R.J.B. Cordero, C.M. de León-Rodriguez, J.K. Alvarado-Torres, A.R. Rodriguez, A. Casadevall, Life Science's Average Publishable Unit (APU) Has Increased over the Past Two Decades, PLOS ONE. 11 (2016) e0156983. https://doi.org/10.1371/journal.pone.0156983.
[6] C. Kim, I. Kozakci, J. Kim, S.Y. Lee, J.-Y. Lee, Highly Efficient (>9%) Lead-Free AgBiS2 Colloidal Nanocrystal/Organic Hybrid Solar Cells, Advanced Energy Materials. 12 (2022) 2200262. https://doi.org/10.1002/aenm.202200262.
[7] Document search - Web of Science Core Collection, (n.d.). https://www-webofscience-com.proxy.bnl.lu/wos/woscc/basic-search (accessed September 12, 2022).
[8] Scopus - Document search, (n.d.). https://www-scopus-com.proxy.bnl.lu/search/form.uri?display=basic#basic (accessed September 12, 2022).
[9] Google Scholar, (n.d.). https://scholar-google-com.proxy.bnl.lu/?pds=12920221543202349033645429711819S (accessed September 12, 2022).

Supplementary information

**Efficiency versus effort: a better way to compare best photovoltaic research cell efficiencies?**

Phillip J. Dale[1] and Michael A. Scarpulla[2]

1. Department of Physics and Materials Science, University of Luxembourg, Belvaux, Luxembourg

2. MSE and ECE Departments, University of Utah, Salt Lake City, Utah, USA

Corresponding authors: phillip.dale@uni.lu and mike.scarpulla@utah.edu

Method

Tabulated below are the search terms used for each technology. These were used in combination with an AND statement containing (photovoltaics OR solar cells). Occasionally a NOT statement is required to remove excessive references from a different technology.

| Technology | NOT |
|---|---|
| CIGS or CIGSe or CuInSe2 or CuGaSe2 or Cu(In,Ga)Se2 or CuInS2 or CuGaS2 or Cu(In,Ga)S2 or CuIn(S,Se)2 or Cu(In,Ga)(S,Se)2 or Cu(In,Al)Se2 | kesterite or Cu2ZnSn* |
| kesterite or Cu2ZnSn* | CIGS or CIGSe or CuInSe2 or CuGaSe2 or Cu(In,Ga)Se2 or CuInS2 or CuGaS2 or Cu(In,Ga)S2 or CuIn(S,Se)2 or Cu(In,Ga)(S,Se)2 or Cu(In,Al)Se2 |
| antimony sulfide or antimony selenide or antimony bismuth selenide or Sb2S3 or Sb2Se3 or Sb2(S,Se)3 or (Sb,Bi)2(Se)3 | |
| Cu2SnS3 or Cu2(Sn,Ge)S3 or Cu2SnSe3 or Cu2Sn(S,Se)3 or Cu2(Sn,Ge)(S,Se)3 | kesterite |
| SnS or tin monosulfide or Sn(S,Se) or SnSxSe1-x | kesterite or CZTS or Cu2ZnSn* |
| CdTe | |
| perovskite or MAPI or CH3NH3PbI3 | |
| Silicon | |
| Quantum dot | |
| Organic or OPV | dye-sensiti*ed OR perovskite |
| dye sensitised or DSSC or dye-sensitized | |
| GaAs | |
| AgBiS2 | |